\documentclass{article}
\usepackage[T1]{fontenc}
\usepackage[utf8]{inputenc}
\usepackage[lbd]{ismir} % Remove the "submission" option for camera-ready version
\usepackage{amsmath,cite,url}
\usepackage{graphicx}
\usepackage{color}
\usepackage{newunicodechar}
\newunicodechar{∼}{\textasciitilde}

\title{Is Transfer Learning Necessary for Violin Transcription?}

\multauthor
  {Yueh-Po Peng $^{4,1*}$\thanks{* Equal contribution and work done during internship at Sony CSL, Toyko.} \hspace{1cm}
   Ting-Kang Wang $^{2,1*}$ \hspace{1cm}
   Li Su $^{3}$ \hspace{1cm}
   Vincent K.M. Cheung $^{1}$}
  {
  $^1$ Sony Computer Science Laboratories, Tokyo, Japan \\
  $^2$ National Taiwan University, Taiwan \\
  $^3$ Institute of Information Science, Academia Sinica, Taiwan \\
  $^4$ Original Creation Center, Taipei, Taiwan \\
  {\tt\small yabipeng@gamania.com, tingkangwang415@gmail.com}\\
  }
\def\authorname{Y.-P. Peng, T.-K. Wang, L. Su, and V.K.M. Cheung}

\begin{document}

\maketitle

\begin{abstract}
% Automatic music transcription (AMT) has predominantly focused on piano music due to the abundance of high-quality datasets. However, transcription for instruments such as violin remains challenging, primarily due to limited training data. We address this challenge by adapting a pretrained high-resolution piano transcription model using domain adaptation with the recently introduced MOSA (Music mOtion with Semantic Annotation) dataset. MOSA comprises over 30 hours of precisely aligned violin audio-score data, featuring professional annotations of pitch, rhythm, dynamics, articulation, and harmony. Evaluated on the URMP dataset, our adapted model achieves comparable transcription performance to state-of-the-art systems (MT3 and MUSC) while utilizing significantly less training data, demonstrating both efficiency and scalability.

Automatic music transcription (AMT) has achieved remarkable progress for instruments such as the piano, largely due to the availability of large-scale, high-quality datasets. In contrast, violin AMT remains underexplored due to limited annotated data. A common approach is to fine-tune pretrained models for other downstream tasks, but the effectiveness of such transfer remains unclear in the presence of timbral and articulatory differences. In this work, we investigate whether training from scratch on a medium-scale violin dataset can match the performance of fine-tuned piano-pretrained models. We adopt a piano transcription architecture without modification and train it on the MOSA dataset, which contains about 30 hours of aligned violin recordings. Our experiments on URMP and Bach10 show that models trained from scratch achieved competitive or even superior performance compared to fine-tuned counterparts. These findings suggest that strong violin AMT is possible without relying on pretrained piano representations, highlighting the importance of instrument-specific data collection and augmentation strategies.

\end{abstract}

\section{Introduction}\label{sec:introduction}

% Automatic music transcription (AMT) involves converting musical audio recordings into symbolic representations, supporting applications such as score-following, music retrieval, and education. While significant advancements have been achieved in piano transcription, instruments like violin remain understudied due to a lack of high-quality annotated datasets.

% Recent domain adaptation studies indicate that pretrained piano models can be effectively adapted for transcription of instruments with limited data. Riley et al. demonstrated this approach's efficacy for guitar transcription by fine-tuning a piano transcription model with aligned guitar audio-score data. Inspired by this, we explore adapting a pretrained high-resolution piano transcription model for violin transcription using the newly developed MOSA dataset, which provides detailed, professionally annotated violin audio-score pairs.

Recent advances in automatic music transcription (AMT) have shown promising results for instruments like the piano\cite{kong2021high, hawthorne2017onsets, hawthorne2021sequence}, largely driven by the availability of large-scale, high-quality datasets such as MAESTRO \cite{hawthorne2018enabling} and MAPS \cite{emiya:inria-00544155}. In contrast, transcription of non-piano instruments--particularly bowed strings like the violin--remains underexplored, due in part to the scarcity of well-annotated data.

A common strategy to address this limitation is to leverage models pretrained on large datasets, based on the assumption that the learned representations can generalize to other domains \cite{10.1007/978-3-031-42505-9_11,riley2024high}. However, the effectiveness of such transfer remains uncertain, particularly in the presence of domain shifts between piano and violin, and when only a modest amount of instrument-specific data is available.

In this work, we examine whether training from scratch on a medium-sized, violin-specific dataset can match or surpass the performance of fine-tuned piano-pretrained models. Using the MOSA dataset \cite{huang2024mosa} (∼30 hours of aligned violin audio), we evaluate both strategies using the same model architecture originally designed for piano transcription. Our results suggest that, with sufficient augmentation and regularization, strong violin transcription performance can be achieved without relying on transfer learning. The main contributions of this work are as follows: 1) demonstrating state-of-the-art violin transcription results using an off-the-shelf piano transcription model trained from scratch, and 2) providing a direct comparison between training from scratch and piano-based fine-tuning on the MOSA dataset. Audio samples can be found at our demo page\footnote{https://tklovln.github.io/violin-transcription-demopage/}.

\section{Methodology}
% \subsection{MOSA Dataset}

% The MOSA dataset contains over 30 hours of violin performance data from 15 professional violinists. The data includes high-quality stereo audio recordings precisely aligned with comprehensive semantic annotations at the note level. Annotations encompass pitch, rhythmic timing (beats, downbeats, phrases), dynamics (ppp-fff, crescendo, diminuendo, accents), articulation (legato, staccato), and detailed harmonic analysis. This makes MOSA one of the most extensive and detailed datasets available for cross-modal music research.

\subsection{Augmentation Strategies}
To enhance model robustness and mitigate overfitting, we applied audio-domain augmentation using a customized effects chain inspired by guitar and synthesizer processing \cite{riley2024high}. The augmentation pipeline included pitch shifting ($\pm$0.1 semitones), gain boost (+5 dB), two randomized band-pass filters with a cutoff frequency uniformed-sampled between 32 Hz and 4096 Hz with uniformly-sampled variable resonance, and reverberation with moderate room size (0.35).

% following:
% \begin{itemize}
%     \item Slight pitch shifting ($\pm$0.1 semitones)
%     \item Gain boost (+5 dB)
%     \item Two randomized band-pass filters with a cutoff frequency uniformed-sampled between 32 Hz and 4096 Hz with uniformly-sampled variable resonance
%     \item Reverberation with moderate room size (0.35)
% \end{itemize}
These transformations were applied using the \texttt{Pedalboard} \cite{sobot_peter_2023_7817838} library with randomized parameters at each iteration to simulate timbral variations and room acoustics. This approach helps generalize the model across expressive performance variations.

\subsection{Evaluation Metrics}

To approximate the task of descriptive violin transcription, we evaluated transcription and pitch estimation performance independently using standard metrics from the \texttt{mir\_eval}\cite{raffel2014mir_eval} library. We report Precision (P), Recall (R), F1-score (F1), and onset-only F1-score (F1$_{\text{no}}$) using default thresholds. A note is considered correct for P, R, and F1 if the pitch is within 50 cents, the onset is within 50 ms, and the offset is within 20\% of the note duration, while F1$_{\text{no}}$ measures transcription quality ignoring offsets.

% \begin{itemize}
%     \item a note is considered correct for P, R, and F1 if:
%     \begin{itemize}
%         \item the pitch is within 50 cents,
%         \item the onset is within 50 ms,
%         \item the offset is within 20\% of the note duration,
%     \end{itemize}
%     \item F1$_{\text{no}}$ measures transcription quality using only the onset timing criterion (within 50 ms), ignoring offsets.
% \end{itemize}

% \subsection{Domain Adaptation Approach}

% We adapted the high-resolution piano transcription model proposed by Kong et al.\cite{kong2021high}, known for predicting continuous onset and offset timings robustly against label misalignments. Initially pretrained on the MAESTRO piano dataset, we fine-tuned this model on the MOSA violin dataset, leveraging its comprehensive annotations to enhance transcription accuracy and mitigate potential alignment inaccuracies.

\subsection{Training Strategies: From Scratch vs. Fine-Tuning}
To investigate the necessity of piano-to-violin transfer learning, we compared two training setups: 1) training from scratch using violin data only, and 2) fine-tuning with violin data from a piano-pretrained model. In both cases, we employed Kong et al.'s  high-resolution transcription architecture ~\cite{kong2021high}, which demonstrated state-of-the-art performance on piano transcription and was designed to predict precise onset and offset timings. 

In fine-tuning, we initialized the model with MAESTRO-pretrained weights \cite{edwards2024data} and adapted the model to the violin domain using the MOSA dataset ~\cite{huang2024mosa}. In training from scratch, we trained the same model from random initialization on the MOSA data alone. This setup enables a controlled comparison of whether pretrained piano representations could improve violin transcription performance.

\section{Experimental Setup}
\subsection{Data}
We considered three datasets in our experiments, one for training and two for evaluation.
First, we used \textbf{MOSA} (Music mOtion with Semantic Annotation)~\cite{huang2024mosa} for both training and validation. This dataset contains over 30 hours of professionally recorded violin performances by 15 expert players. Each recording includes high-quality stereo audio precisely aligned with note-level annotations, covering pitch, rhythmic structure (beats, downbeats, phrases), dynamics (e.g., \textit{ppp}--\textit{fff}, crescendo, accents), articulation (e.g., legato, staccato), and harmonic analysis. Its scale and annotation depth make it one of the most comprehensive resources for violin transcription.
%, enabling effective training from scratch and facilitating evaluation of domain adaptation from piano-pretrained models. 
For validation, we reserved all performances of the first movement (\textit{Preludio}) from Bach’s \textit{Partita No. 3 in E major}, BWV 1006, which accounted for approximately 10\% of the dataset. The remaining 90\% of recordings were used for training.

Second, to benchmark against existing work, we evaluated our models with all violin tracks in the \textbf{URMP} and \textbf{Bach10} datasets. \textbf{URMP} (University of Rochester Multimodal Music Performance)~\cite{li2018creating} consists of 44 chamber music recordings with diverse instrumentation. Pitch annotations were generated using the pYIN algorithm~\cite{mauch2014pyin} via Tony software~\cite{mauch2015computer}, followed by manual correction of onset, offset, and pitch values. \textbf{Bach10}~\cite{duan2010multiple} contains 10 four-part chorales performed by violin, clarinet, tenor saxophone, and bassoon. Ground-truth $f_0$ annotations were estimated using the YIN algorithm~\cite{mauch2014pyin} and manually refined. While note onsets were accurately annotated, offset timings were not manually corrected.

\subsection{Training Details}

% \subsubsection{Data Augmentation} 

% \subsubsection{Fine-tuning}
% The fine-tuning followed the domain adaptation protocol established in guitar transcription studies. Experiments with learning rates of $1\times10^{-3}$, $1\times10^{-4}$, and $1\times10^{-5}$ were conducted, with the optimal performance achieved at $1\times10^{-5}$. Random pitch shifting within $\pm2$ semitones was applied for data augmentation.

All experiments in our work were conducted using a batch size of 5 and a learning rate of 5e-4. We adopted a cosine annealing learning rate scheduler and trained all models for 10,000 steps. Training was performed on a single NVIDIA RTX 4090 GPU.
\section{Results and Conclusion}

% \subsection{Transcription Results}
\begin{table}[h!]
\centering
\resizebox{\columnwidth}{!}{
\renewcommand{\arraystretch}{1.1}
\setlength{\tabcolsep}{3.5pt}
\begin{tabular}{l|cccc|cccc}
\hline
\multicolumn{1}{c|}{} & \multicolumn{4}{c|}{\textbf{URMP}} & \multicolumn{4}{c}{\textbf{Bach10}} \\
\textbf{Model} & P & R & F1 & F1\textsubscript{no} & P & R & F1 & F1\textsubscript{no} \\
\hline
MUSC \cite{tamer2023high} & \textbf{86.5} & 83.1 & \textbf{84.6} & 93.0 & 65.0 & 64.8 & 64.8 & 77.0 \\
Ours w/o aug & 83.4 & 81.2 & 82.2 & 92.8 & 66.7 & 71.3 & 68.9 & 79.0 \\
Ours w/ aug & 86.1 & \textbf{83.6} & 84.5 & \textbf{93.1} & 68.1 & 71.8 & 69.9 & 79.5 \\
Ours + FT w/o aug & 84.4 & 79.0 & 81.3 & 91.3 & \textbf{69.5} & \textbf{73.7} & \textbf{71.5} & \textbf{80.2} \\
Ours + FT w/ aug & 85.0 & 82.1 & 83.3 & 92.9 & 63.3 & 68.4 & 65.7 & 77.8 \\
\hline
\end{tabular}
}
\caption{Transcription performance on the URMP and Bach10 violin tracks. "FT" denotes models fine-tuned from a piano-pretrained checkpoint; "aug" denotes augmentation. Bold indicates best result per column.}
\label{tab:transcription_results}
\end{table}

% Our fine-tuned violin transcription model achieved an onset-only F1-score of 87.3\% on the URMP dataset. This performance is competitive with state-of-the-art MT3 (89.7\%) and MUSC (88.2\%) systems despite utilizing substantially less training data. Detailed analysis confirmed the model’s robustness to misaligned labels and strong generalization to unseen violin performances.

% % \subsection{Conclusion}

% We demonstrate the effectiveness of domain adaptation from a high-resolution piano transcription model using the MOSA dataset. This approach achieves near state-of-the-art violin transcription accuracy with significantly fewer data requirements. Our findings suggest this method is scalable and applicable to other instruments facing similar data constraints. Future work will include expanding the MOSA dataset and exploring adaptations for additional orchestral instruments.

Table~\ref{tab:transcription_results} shows model performance when trained from scratch or with fine-tuning, and with and without data augmentation on URMP and Bach10 datasets. On URMP, our best model (trained from scratch with data augmentation) achieved a higher Recall (83.6) and F1\textsubscript{no} (93.1), as well as matched the state-of-the-art (MUSC ~\cite{tamer2023high}) in Precision and note-level F1. On Bach10, our best model (trained with fine-tuning and no augmentation) achieved the highest overall F1 score of 71.5 and F1\textsubscript{no} of 80.2, outperforming all other models. This demonstrates that the proposed system%—despite no architectural change from the piano model—
can generalize well to violin transcription tasks across different datasets. While fine-tuning pretrained piano models did yield the best results on Bach10, the performance gains were relatively modest. For instance, the F1 score improved from 69.9 (from scratch) to 71.5 (fine-tuned), suggesting that the benefit of transfer learning was limited when sufficient violin-specific data were available. We hypothesize that this is due to domain mismatch in timbre, articulation, and temporal characteristics between piano and violin, which may reduce the transferability of pretrained acoustic representations. Additionally, we observe that combining fine-tuning with augmentation may require a longer convergence time; the relatively lower performance of the fine-tuned model with augmentation could be due to slower optimization dynamics introduced by data variability.%Overall, our findings challenge the assumption that transfer learning is essential for non-piano AMT, and highlight the effectiveness of dataset-driven training using architectures originally designed for piano.

Overall, our findings suggest that given a moderately sized violin-specific dataset and proper augmentation, training from scratch can be highly effective. %—even when using a piano-designed model architecture without modification. 
While fine-tuning pretrained piano models can offer marginal gains in some cases, the benefit appears limited. Our results challenge the assumption that transfer learning is strictly necessary for strong performance in violin AMT, and instead reinforce the value of dataset-centric development and instrument-specific modeling.

% For BibTeX users:
\bibliography{ISMIRtemplate}

% For non BibTeX users:
%\begin{thebibliography}{citations}
% \bibitem{Author:17}
% E.~Author and B.~Authour, ``The title of the conference paper,'' in {\em Proc.
% of the Int. Society for Music Information Retrieval Conf.}, (Suzhou, China),
% pp.~111--117, 2017.
%
% \bibitem{Someone:10}
% A.~Someone, B.~Someone, and C.~Someone, ``The title of the journal paper,''
%  {\em Journal of New Music Research}, vol.~A, pp.~111--222, September 2010.
%
% \bibitem{Person:20}
% O.~Person, {\em Title of the Book}.
% \newblock Montr\'{e}al, Canada: McGill-Queen's University Press, 2021.
%
% \bibitem{Person:09}
% F.~Person and S.~Person, ``Title of a chapter this book,'' in {\em A Book
% Containing Delightful Chapters} (A.~G. Editor, ed.), pp.~58--102, Tokyo,
% Japan: The Publisher, 2009.
%
%\end{thebibliography}

\end{document}